\documentclass[12pt,a4paper]{article}
\usepackage{amsmath,amssymb,slashed,cancel}
\usepackage{cite,tabularx}
\usepackage{subcaption}
\usepackage{graphicx,color}
\usepackage[normalem]{ulem}
\usepackage{mathtools}
\usepackage{enumerate}
\usepackage{multirow}
\usepackage{bm}
\usepackage{braket}
\usepackage{arydshln}
\usepackage[subrefformat=parens]{subcaption}
\allowdisplaybreaks

\usepackage[height=8.85in,width=6.4in]{geometry}

\newcommand{\GeV}{\,\mathrm{GeV}}

\newcommand{\DM}{\mathrm{DM}}

\newcommand\package[2][\relax]{\texttt{#2\ifx#1\relax\relax\relax\else\,\linebreak[0]#1\fi}}

\numberwithin{equation}{section} 
 
\def\beq#1\eeq{\begin{align}#1\end{align}}

\definecolor{BlueViolet}{rgb}{0.2, 0.00, 0.7}
\definecolor{Blue}{rgb}{0.15, 0.00, 0.9}
\usepackage[colorlinks=true,linkcolor=Blue,citecolor=Blue,urlcolor=BlueViolet]{hyperref}

\begin{document}
\begin{titlepage}
\setcounter{page}{0} 

\begin{flushright}
KYUSHU-HET-241
\end{flushright}

\begin{center}

\vskip .55in

\begingroup
\centering
\large\bf 
Scalar Dark Matter with a $\mu\tau$ Flavored Mediator

\endgroup

\vskip .4in

\renewcommand{\thefootnote}{\fnsymbol{footnote}}
{
Kento Asai$^{(a,b)}$\footnote{
  \href{mailto:kento@icrr.u-tokyo.ac.jp}
  {\tt kento@icrr.u-tokyo.ac.jp}},
Coh Miyao$^{(c)}$\footnote{
  \href{mailto:miyao.coh@phys.kyushu-u.ac.jp}
  {\tt miyao.coh@phys.kyushu-u.ac.jp}},
Shohei Okawa$^{(d)}$\footnote{
  \href{mailto:okawa@icc.ub.edu}
  {\tt okawa@icc.ub.edu}}, and
Koji Tsumura$^{(c)}$\footnote{
  \href{mailto:tsumura.koji@phys.kyushu-u.ac.jp}
  {\tt tsumura.koji@phys.kyushu-u.ac.jp}}
}

\vskip 0.4in

\begingroup\small
\begin{minipage}[t]{1.0\textwidth}
\centering\renewcommand{\arraystretch}{1.0}
{\it
\begin{tabular}{c@{\,}l}
$^{(a)}$
& Institute for Cosmic Ray Research (ICRR), The University of Tokyo, Kashiwa, \\
& Chiba 277--8582, Japan \\[2mm]
$^{(b)}$
& Department of Physics, Faculty of Engineering Science, Yokohama National University, \\
& Yokohama 240--8501, Japan \\[2mm]
$^{(c)}$
& Department of Physics, Kyushu University, 744 Motooka, Nishi-ku, Fukuoka 819--0395, \\
& Japan \\[2mm]
$^{(d)}$
& Departament de F\'{i}sica Qu\`{a}ntica i Astrof\'{i}sica, Institut de Ci\`{e}ncies del Cosmos (ICCUB), \\
& Universitat de Barcelona, Mart\'{i} i Franqu\`{e}s 1, E-08028 Barcelona, Spain \\[2mm]
\end{tabular}
}
\end{minipage}
\endgroup

\end{center}

\vskip .4in

\begin{abstract}\noindent
We study a renormalizable scalar singlet dark matter model based on $Z_4$ lepton flavor symmetry.
A $\mu\tau\mathchar`-$philic scalar doublet is introduced as a mediator which connects dark matter and standard model particles. 
The observed relic abundance of the dark matter is easily maintained while satisfying the current severe constraints on the dark matter from various experiments and observations thanks to the flavor off-diagonal interactions of scalar mediators. 
We further explore the possibility of dark matter direct detection through the one-loop process.
We also find that the relic abundance of the dark matter and the discrepancy of the muon anomalous magnetic moment can be explained by the $\mu\tau\mathchar`-$philic scalar mediator simultaneously.
\end{abstract}
\end{titlepage}

\setcounter{page}{1}
\renewcommand{\thefootnote}{\#\arabic{footnote}}
\setcounter{footnote}{0}


\section{Introduction}
\label{sec:introduction}

The Standard Model (SM) in particle physics has been established by the discovery of the Higgs boson at the LHC. 
It successfully explains a variety of experimental results thus far.
In the meantime, a lot of experiments have been undertaken around the world 
to verify the SM with unprecedented accuracy or discover new phenomena induced by physics beyond the SM. 
Among the past experimental results, 
the precise measurement of the muon anomalous magnetic moment (muon $g-2$) is of particular interest. 
The measured value of the muon $g-2$ at Brookhaven National Laboratory~\cite{Muong-2:2002wip,Muong-2:2004fok,Muong-2:2006rrc} and
at Fermilab~\cite{Muong-2:2021ojo} 
indicates the discrepancy between the theoretical and experimental values
\begin{align}
\label{eq:muong-2_exp}
   \Delta a_\mu = a_\mu^{\rm exp} - a_\mu^{\rm SM} = ( 25.1 \pm 5.9) \times 10^{-10}~,
\end{align}
whose significance is about to $4.2\sigma$.
Moreover, the upcoming experiment at J-PARC based on a different technique
by using ultra-cold muons~\cite{Mibe:2011zz} will provide independent information as to the systematic uncertainties. 
The discrepancy, if confirmed, reveals the existence of new physics which couples to the SM lepton sector. 

With this long-standing problem,  
a lot of new physics models related to the lepton sector are proposed in the last decade.  
One direction in model building is the extension of the lepton flavor sector by a gauge symmetry. 
For example, the difference between the muon and tau flavor numbers is gauged as U(1)$_{L_\mu-L_\tau}$ symmetry~\cite{Foot:1990mn,He:1990pn,He:1991qd,Foot:1994vd}.
The associated gauge boson does not couple to the electron and quarks at the tree level~\cite{Baek:2001kca,Ma:2001md,Heeck:2011wj,Harigaya:2013twa}. 
On the other hand, the new gauge boson induces a sizable contribution to the muon $g-2$ without conflicting 
existing constraints from collider experiments and cosmological and astrophysical observations.

As another possibility, there is a class of extension of the lepton flavor sector by discrete symmetry. 
In its minimal model based on a $Z_4$ symmetry~\cite{Abe:2019bkf}, there is an additional scalar doublet 
which only has a flavor off-diagonal coupling to the second and third generation leptons 
because of the discrete flavor symmetry\footnote{A similar Yukawa interaction can also be realized by a gauged SU(2)${}_{\mu\tau}$ symmetry~\cite{Chiang:2017vcl}.}. 
This $\mu\tau$ flavored interaction induces the tau mass enhanced contribution to the muon $g-2$, 
thereby allowing to resolve the discrepancy. 

Dark matter (DM) is another puzzle in cosmology.
There is no candidate for DM in the SM, and therefore, various types of DM have been proposed.
An attractive candidate of DM is thermal relic of stable massive particles, which are produced through their interactions to SM particles in the thermal bath in the early Universe. 
The cosmological abundance of the thermal relic DM is determined by its annihilation cross section into SM particles.
In the thermal relic scenario, interactions between the DM and SM particles, which are responsible for the production, also cause non-negligible scattering of DM with nucleons and electrons in general.
In spite of considerable efforts to directly detect such a scattering process, however, no positive signal has been observed so far.
As a result, DM interactions with the SM particles, especially quarks and electrons, are strictly constrained, motivating model building efforts to sufficiently suppress such DM interactions.

In this paper, we reinforce the $Z_4$ model in Ref.~\cite{Abe:2019bkf} with a $\mu\tau$ flavored complex scalar singlet DM\footnote{
Models with gauged $\mu\tau$ flavored DM are shown in Refs.~\cite{Baek:2008nz,Foldenauer:2018zrz,Garani:2019fpa,Asai:2020qlp,Holst:2021lzm}.
} and 
explore the possibility that the $\mu\tau\mathchar`-$philic scalar doublet acts as a mediator between the dark sector and the SM sector.
The relic abundance of DM is thermally produced through the interaction to the $\mu\tau\mathchar`-$philic scalar.
We show that the relic abundance is correctly produced in a broad mass range of such a DM candidate.
A remarkable feature in this model is that due to the $Z_4$ flavor symmetry, the $\mu\tau\mathchar`-$philic scalar mediator only has flavor off-diagonal couplings to the muon and tau flavor leptons, which forbids tree-level couplings with the quarks and electrons.
The flavor symmetry also guarantees the DM stability.
DM and mediators with lepton flavor off-diagonal couplings are studied in Ref.~\cite{Galon:2016bka} in an effective field theory framework. 
The model studied in the present paper can thus be regarded as a renormalizable example.
It turns out that scattering processes relevant to DM direct detection are absent at the tree level thanks to the lepton flavor symmetry, 
while the one-loop $Z$ boson exchange process can generate DM-nucleon scattering at a detectable rate, which is in sharp contract to the studies in Ref.~\cite{Galon:2016bka}. 
Furthermore, indirect gamma-ray searches at the Fermi-LAT experiment~\cite{Hoof:2018hyn} 
exclude the low mass region where the direct detection hardly operates. 
This complementarity restricts the viable mass range to be above a few tens GeV.
We also find that the addition of the DM candidate does not spoil the success in explaining the discrepancy of the muon $g-2$ by the $\mu\tau\mathchar`-$philic scalar mediator, which was pointed out in Ref.~\cite{Abe:2019bkf}.

This paper is organized as follows. 
In Sec.~\ref{sec:model}, we introduce the model involving a $\mu\tau\mathchar`-$philic scalar doublet and a complex scalar DM. 
The contribution of the $\mu\tau\mathchar`-$philic scalars to the muon $g-2$ is also shown there.
We discuss the DM phenomenology in Sec.~\ref{sec:DM} and show the theoretical and experimental constraints on the $\mu\tau\mathchar`-$philic mediator in Sec.~\ref{sec:mediator}.
We show our result in Sec.~\ref{sec:result} and, the summary is finally given in Sec.~\ref{sec:summary}.

\section{Model}
\label{sec:model}
In addition to the SM gauge symmetry, we introduce an Abelian discrete symmetry $Z_4$ under which 
muon and tau flavor leptons transform nontrivially and the other SM particles are singlet. See Tab.~\ref{tab:fields} for the particle contents and charge assignment.
\begin{table}[tb]
    \centering
    \begin{tabular}{|c|ccccc|} \hline
         particles & $(L_e, L_\mu, L_\tau)$ & $(e_R^{}, \mu_R^{}, \tau_R^{})$ & $H$ & $\Phi$ & $\Sigma$  \\ \hline
         SM & $(\bm{1},\bm{2})_{-1/2}$ & $(\bm{1},\bm{1})_{-1}$ & $(\bm{1},\bm{2})_{1/2}$ & $(\bm{1},\bm{2})_{1/2}$ & $(\bm{1},\bm{1})_{0}$ \\
         $Z_4$ & $(1, i, -i)$ & $(1, i, -i)$ & $1$ & $-1$ & $i$ \\ \hline
    \end{tabular}
    \caption{
    Particle contents of the scalar DM model with $\mu\tau\mathchar`-$philic mediator.
    The quantum numbers of the SM are also shown in the notation of $({\rm SU(3)}_C, {\rm SU(2)}_L)_{{\rm U(1)}_Y}$.
    }
    \label{tab:fields}
\end{table}
Apart from them, we introduce a SM gauge singlet complex scalar $\Sigma$ and SU(2)$_L$ doublet scalar $\Phi$.
These scalars play roles of DM and $\mu\tau$-philic mediator, respectively.
Under the SM gauge and discrete flavor symmetries, the following Lagrangian is allowed,
\begin{align}
\label{eq:Lagrangian}
   \mathcal{L} 
   &= \mathcal{L}_{\rm SM} 
      + |D_\mu \Phi|^2 + |D_\mu \Sigma|^2
      - \left( y_{\mu\tau} L_\mu^\dag \Phi \tau_R^{} + y_{\tau\mu} L_\tau^\dag \Phi \mu_R^{}  + \text{H.c.} \right)
      - V(H,\Phi,\Sigma)~, \\
\label{eq:potential}
   V(H,\Phi,\Sigma) 
   &= \mu_\Phi^2 |\Phi|^2 + \lambda_2 |\Phi|^4 +\lambda_3 |H|^2 |\Phi|^2 + \lambda_4 |H^\dag \Phi|^2 + \frac{\lambda_5}{2} \left[ (H^\dag \Phi)^2 + \text{H.c.} \right] \nonumber \\
   &~~~ + \mu_\Sigma^2 |\Sigma|^2 + \lambda_\Sigma^{} |\Sigma|^4 + [\lambda'_\Sigma \Sigma^4 + \text{H.c.}] + \lambda_{H\Sigma}^{} |H|^2 |\Sigma|^2 + \lambda_{\Phi\Sigma}^{} |\Phi|^2 |\Sigma|^2 \nonumber \\
   &~~~ + \kappa \left[ (H^\dag \Phi) \Sigma^2 + \text{H.c.} \right]~,
\end{align}
where the first term of Eq.~\eqref{eq:Lagrangian} stands for the Lagrangian of the SM.
The $\mu\tau$-philic mediator $\Phi$ has a flavor off-diagonal Yukawa coupling between muon and tau lepton. 
In general, the Yukawa couplings $y_{\mu\tau/\tau\mu}$ and scalar quartic couplings $\lambda'_\Sigma$, $\lambda_5, \kappa$ are complex. 
Among them, two phases of the scalar quartic couplings can be removed by the phase redefinition of $\Phi$ and $\Sigma$. 
In this paper, therefore, we rotate away the phases of $\lambda_5$ and $\kappa$ and choose them to be real. 
Moreover, we also assume that the $\mu\tau$-philic mediator $\Phi$ and a singlet scalar $\Sigma$ have no vacuum expectation value (VEV), and therefore $Z_4$ is exact. 
The SM Higgs doublet $H$ acquires a non-zero VEV as usual and breaks electroweak symmetry. 
The DM stability is guaranteed by the accidental $Z_2$ symmetry, $\Sigma \to -\Sigma$, which is 
realized as a part of the $Z_4$ symmetry~\footnote{
For the successful generation of the neutrino masses and mixing, the $Z_4$ symmetry has to be broken completely.
In this case, an additional symmetry such as dark parity is needed for the DM stability.
One of the extended model which realizes the neutrino masses and mixing is shown in Appendix~\ref{app:neutrino}.
This kind of extension does not much affect the DM phenomenology, and therefore we concentrate on the above simple model in this paper.
}.

The scalar doublet fields can then be parameterized as
\begin{align}
    H 
    = \left( 0, (v + h)/\sqrt{2} \right)^T~,~~~~~
    \Phi
    = \left( \phi^+, (\rho + i \eta)/\sqrt{2} \right)^T~,
\end{align}
where $\phi^+$, $\rho$, and $\eta$ stand for an electrically charged, and two neutral scalars, respectively, and $v = 246$\,GeV is the VEV of the SM Higgs boson $h$.
For the above potential, the masses of new scalar particles are given by
\begin{align}
    m_{\phi^\pm}^2 
    &= \mu_\Phi^2 + \frac{1}{2} \lambda_3 v^2~, \\
    m_\rho^2
    &= \mu_\Phi^2 + \frac{1}{2} (\lambda_3 + \lambda_4 + \lambda_5) v^2~, \\
    m_\eta^2
    &= \mu_\Phi^2 + \frac{1}{2} (\lambda_3 + \lambda_4 - \lambda_5) v^2~, \\
    m_\Sigma^2
    &= \mu_\Sigma^2 + \frac{1}{2} \lambda_{H\Sigma}^{} v^2~.
\end{align}
Note that neutral mediators $\rho$ and $\eta$ are real fields, 
while the DM candidate scalar $\Sigma$ behaves as a complex field.

The contributions to the muon $g-2$ from $\rho$ and $\eta$ are calculated by~\cite{Abe:2019bkf}
\begin{align}
\label{eq:muong-2}
   \Delta a_\mu^{\rm new}
   &= \frac{{\rm Re}(y_{\mu\tau} y_{\tau\mu})}{(4\pi)^2} \left[ \frac{m_\mu m_\tau}{m_\rho^2} I_1(m_\mu^2/m_\rho^2, m_\tau^2/m_\rho^2) - \frac{m_\mu m_\tau}{m_\eta^2} I_1(m_\mu^2/m_\eta^2, m_\tau^2/m_\eta^2) \right] \nonumber \\
   &~~~ + \frac{|y_{\mu\tau}|^2 + |y_{\tau\mu}|^2}{2 (4\pi)^2} \left[ \frac{m_\mu^2}{m_\rho^2} I_2(m_\mu^2/m_\rho^2, m_\tau^2/m_\rho^2) + \frac{m_\mu^2}{m_\eta^2} I_2(m_\mu^2/m_\eta^2, m_\tau^2/m_\eta^2) \right]~,
\end{align}
where the loop functions $I_1(\alpha, \beta)$ and $I_2(\alpha, \beta)$ are defined by
\begin{align}
\label{eq:loop-func}
   I_1(\alpha, \beta) 
   &\equiv \int_0^1 {\rm d}x \frac{(1-x)^2}{x - x (1-x) \alpha + (1-x) \beta}~, \\
   I_2(\alpha, \beta) 
   &\equiv \frac{1}{2} \int_0^1 {\rm d}x \frac{x (1-x)^2}{x - x (1-x) \alpha + (1-x) \beta}~.
\end{align}
When both $y_{\mu\tau}$ and $y_{\tau\mu}$ are nonvanishing, 
the dominant contribution arises from the first line of Eq.~\eqref{eq:muong-2} since it enjoys 
the enhancement by a factor of $m_\tau/m_\mu$ compared with that from the second line.
Note also that the tau mass enhanced contribution is asymmetric under the exchange $\rho \leftrightarrow \eta$. 
As a result, the sign of the muon $g-2$ contribution is thus determined by the sign of $(m_\eta - m_\rho) \times {\rm Re}(y_{\mu\tau} y_{\tau\mu})$. 
The appropriate sign to diminish the discrepancy is obtained when $(m_\eta - m_\rho) \times {\rm Re}(y_{\mu\tau} y_{\tau\mu}) > 0$.  
In the following discussion, we assume that the parameters in this model are fixed to accommodate the muon $g-2$ anomaly within $2\sigma$.

\section{Dark Matter Physics}
\label{sec:DM}

In this section, we study the DM thermal production and DM direct/indirect detections.

\subsection{Relic Density}
\label{subsec:relic}

We assume the thermal freeze-out scenario as the DM production in this paper. 
Thus, the relic density of the DM is determined by the DM annihilation cross section.
In this model, the DM annihilates through two types of processes, namely the $\mu\tau$-philic scalar and Higgs mediated ones.
The thermal freeze-out scenario through the pure Higgs mediated process (so-called Higgs portal scenario) is exhaustively studied in various new physics models. The current constraints on this scenario at the direct detection experiments restrict its allowed region to the Higgs resonance or heavy mass region above a few TeV. 
Therefore, we simply assume $\lambda_{H\Sigma}^{} = 0$ in this paper and focus only on the annihilation mediated by the $\mu\tau$-philic scalars.

\begin{figure}
    \centering
    \includegraphics[viewport=200 600 400 770, clip=true, scale=0.65]{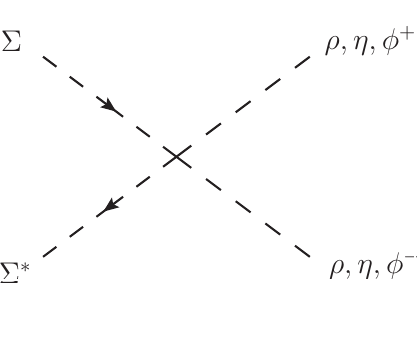}
    \hspace{1cm}
    \includegraphics[viewport=180 560 400 770, clip=true, scale=0.6]{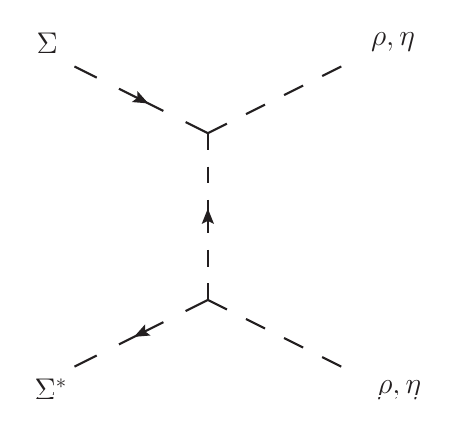}
    \caption{
    Feynman diagrams for $\Sigma \Sigma^*$ annihilation. 
    The arrow in the diagrams represents the flow of the $\Sigma$ number current. 
    }
    \label{fig:SigSigbar}
\end{figure}

\begin{figure}
    \centering
    \hspace{.5cm}
    \includegraphics[viewport=200 628 380 770, clip=true, scale=0.65]{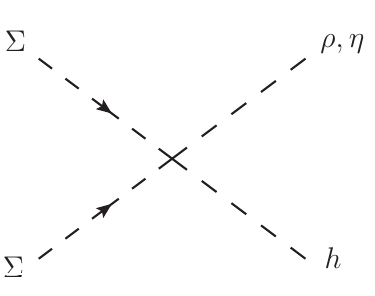}
    \hspace{1.3cm}
    \includegraphics[viewport=170 620 440 770, clip=true, scale=0.6]{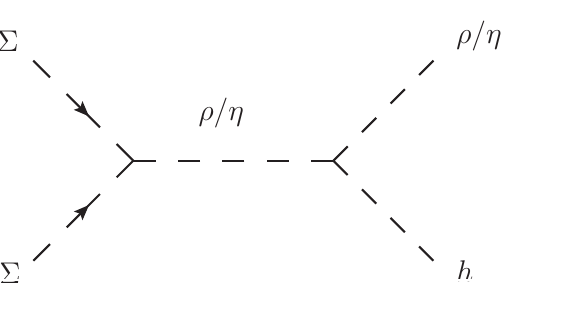}\\
    \includegraphics[viewport=170 620 440 770, clip=true, scale=0.6]{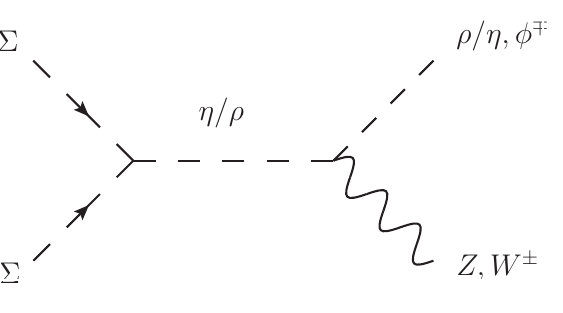}
    \hspace{.1cm}
    \includegraphics[viewport=170 615 430 770, clip=true, scale=0.6]{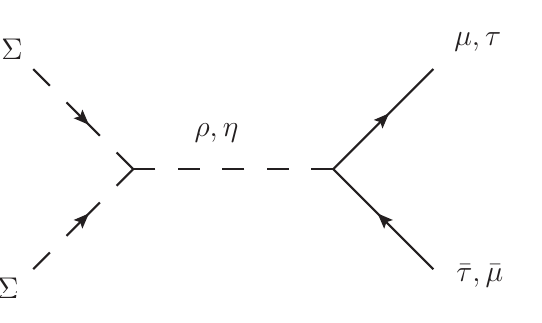}
    \caption{
    Feynman diagrams for $\Sigma \Sigma$ annihilation.
    $\Sigma^* \Sigma^*$ annihilation arises via similar diagrams.
    }
    \label{fig:SigSig}
\end{figure}

The DM $\Sigma$ can annihilate with its antiparticle or itself.
For the former case, the DM pairs $(\Sigma\Sigma^*)$ annihilate into $\phi^+\phi^-, \rho\rho, \eta\eta,$ and $\rho\eta$ through two kinds of diagrams in Fig.~\ref{fig:SigSigbar}.
The cross sections of the annihilation of the DM and its antiparticle are given by, 
at the leading order in the expansion of the DM relative velocity $v_{\rm rel}$, 
\begin{align}
\label{eq:xann-SigSigbar}
   (\sigma v_{\rm rel})_{\Sigma \Sigma^* \to \phi^+ \phi^-} 
   &= \frac{\lambda_{\Phi\Sigma}^2}{32\pi m_{\Sigma}^2} \sqrt{1-\frac{m_\phi^2}{m_\Sigma^2}}~, \\
   (\sigma v_{\rm rel})_{\Sigma \Sigma^* \to \rho \rho} 
   &= \frac{1}{64\pi m_\Sigma^2} \left( \frac{2 \kappa^2 v^2}{2m_\Sigma^2 - m_\rho^2} - \lambda_{\Phi\Sigma}^{} \right)^2 \sqrt{1-\frac{m_\rho^2}{m_\Sigma^2}}~, \\
   (\sigma v_{\rm rel})_{\Sigma \Sigma^* \to \eta \eta} 
   &= (\sigma v_{\rm rel})_{\Sigma \Sigma^* \to \rho \rho} \bigg|_{m_\rho \to m_\eta}~, \\
   (\sigma v_{\rm rel})_{\Sigma \Sigma^* \to \rho \eta} 
   &= \frac{2 v_{\rm rel}^2}{3\pi} \frac{(\kappa v)^4 m_\Sigma^2}{(4m_\Sigma^2 - m_\rho^2 - m_\eta^2)^4} \beta_{\rho\eta}^3 
\end{align}
with 
\begin{align}
   \beta_{ij} 
   = \sqrt{1 - \frac{m_i^2 + m_j^2}{2m_\Sigma^2} + \frac{(m_i^2-m_j^2)^2}{(4m_\Sigma^2)^2} }~.
\end{align}
Note that $\Sigma \Sigma^* \to \rho \eta$ is $p$-wave process. 
For the latter case, a pair of the DM particles $(\Sigma\Sigma)$ annihilates into 
$\rho h, \rho Z, \eta h, \eta Z, \phi^- W^+, \phi^+ W^-, \mu^+ \tau^-,$ and $\mu^- \tau^+$, 
as shown by the Feynman diagrams in Fig.~\ref{fig:SigSig}.
The cross sections of these annihilation processes are given by
\begin{align}
\label{eq:xann-SigSig}
   (\sigma v_{\rm rel})_{\Sigma \Sigma \to \rho h} 
   &= \frac{\kappa^2}{32\pi m_\Sigma^2} \left( 1 + \frac{v^2 (\lambda_3 + \lambda_4 + \lambda_5)}{4 m_\Sigma^2 - m_\rho^2} \right)^2 \beta_{\rho h}~, \\
   (\sigma v_{\rm rel})_{\Sigma \Sigma \to \rho Z} 
   &= \frac{\kappa^2}{32\pi m_\Sigma^2} \left(\frac{4m_\Sigma^2}{4m_\Sigma^2-m_\eta^2}\right)^2 \beta_{\rho Z}^3~, \\
   (\sigma v_{\rm rel})_{\Sigma \Sigma \to \eta h, \eta Z} 
   &= (\sigma v_{\rm rel})_{\Sigma \Sigma \to \rho h, \rho Z} \bigg|_{\substack{m_\rho \leftrightarrow m_\eta \\ \lambda_5 \to -\lambda_5}}~, \\
   (\sigma v_{\rm rel})_{\Sigma \Sigma \to \phi^- W^+} 
   &= \frac{\kappa^2}{32\pi m_\Sigma^2} \left( \frac{4m_\Sigma^2}{4m_\Sigma^2-m_\rho^2} + \frac{4m_\Sigma^2}{4m_\Sigma^2-m_\eta^2} \right)^2 \beta_{\phi W}^3~, \\
   (\sigma v_{\rm rel})_{\Sigma \Sigma \to \phi^+ W^-} 
   &= \frac{\kappa^2}{32\pi m_\Sigma^2} \left( \frac{4m_\Sigma^2}{4m_\Sigma^2-m_\rho^2} - \frac{4m_\Sigma^2}{4m_\Sigma^2-m_\eta^2} \right)^2 \beta_{\phi W}^3~, \\
   (\sigma v_{\rm rel})_{\Sigma \Sigma \to \mu \bar{\tau}} 
   &= \frac{\kappa^2 v^2}{64\pi m_\Sigma^2} \left[ a \, |y_{\mu\tau}|^2 + b \, |y_{\tau\mu}|^2 + c \, {\rm Re}(y_{\mu\tau} y_{\tau\mu} ) \right] \beta_{\mu\tau}~, \\
   (\sigma v_{\rm rel})_{\Sigma \Sigma \to \tau \bar{\mu}} 
   &= \frac{\kappa^2 v^2}{64\pi m_\Sigma^2} \left[ b \, |y_{\mu\tau}|^2 + a \, |y_{\tau\mu}|^2 + c \, {\rm Re}(y_{\mu\tau} y_{\tau\mu} ) \right] \beta_{\mu\tau}
\end{align}
with 
\begin{align}
   a 
   &= (4m_\Sigma^2 - m_\mu^2 - m_\tau^2) \left( \frac{1}{4m_\Sigma^2 - m_\rho^2} - \frac{1}{4m_\Sigma^2 - m_\eta^2} \right)^2~, \\
   b 
   &= (4m_\Sigma^2 - m_\mu^2 - m_\tau^2) \left( \frac{1}{4m_\Sigma^2 - m_\rho^2} + \frac{1}{4m_\Sigma^2 - m_\eta^2} \right)^2~,\\
   c 
   &= -4m_\mu m_\tau \left[ \left( \frac{1}{4m_\Sigma^2 - m_\rho^2} \right)^2 - \left( \frac{1}{4m_\Sigma^2 - m_\eta^2} \right)^2 \right]~.
\end{align}
It should be noted that in the resonant regime ($m_\Sigma^{} \simeq m_{\rho,\eta}/2$), we have to properly replace the $s$-channel propagators with the Breit-Wigner form.

The relic abundance of the DM is calculated by solving the Boltzmann equation, 
\begin{align}
\label{eq:Boltzmann}
   \frac{{\rm d}n_{\rm DM}^{}}{{\rm d}t} + 3 H n_{\rm DM}^{}
   &= - \frac{1}{2} (\sigma v_{\rm rel})_{\rm eff} \left[ n_{\rm DM}^2 - (n_{\rm DM}^{\rm eq})^2 \right]~,
\end{align}
where $H$ is the Hubble parameter, and $n_{\rm DM}^{\rm (eq)} = n_{\Sigma}^{\rm (eq)} + n_{\Sigma^*}^{\rm (eq)}$ stands for the DM total (equilibrium) number density. 
The effective annihilation cross section is defined by 
\begin{align}
   (\sigma v_{\rm rel})_{\rm eff}
   &= \sum_{ij} \left[ (\sigma v_{\rm rel})_{\Sigma \Sigma^* \to ij} + \frac{1}{2} (\sigma v_{\rm rel})_{\Sigma \Sigma \to ij} + \frac{1}{2} (\sigma v_{\rm rel})_{\Sigma^* \Sigma^* \to ij} \right]~.
\end{align}
The approximate formula for the produced DM density parameter is well known and given in terms of the effective annihilation cross section,
\begin{align}
\label{eq:relic-approx}
   \Omega_{\rm DM}^{} h^2
   \simeq 0.12 \times \left( \frac{3 \times 10^{-26}~{\rm cm^3/s}}{\frac{1}{2}(\sigma v_{\rm rel})_{\rm eff}} \right)~,
\end{align}
where $h$ stands for the renormalized Hubble parameter, which is related to the present value of 
the Hubble parameter as $H_0 = 100\, h~ {\rm km\, sec^{-1}\, Mpc^{-1}}$.
In our analysis, the relic abundance of the DM is numerically calculated by \texttt{micrOMEGAs\_5\_2\_4}~\cite{Belanger:2018ccd} instead of using Eq.~\eqref{eq:relic-approx}, and the model file is generated by \texttt{Feynrules\_2\_3}~\cite{Alloul:2013bka}.

\subsection{Direct Detection}
\label{subsec:direct}

\begin{figure}[t]
\centerline{
\includegraphics[viewport=160 580 420 760, clip=true, scale=0.7]{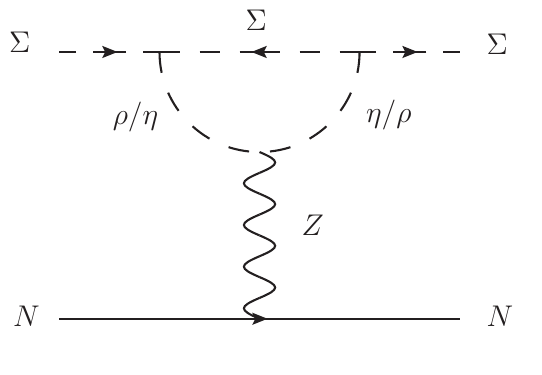}
}
\caption{$Z$ penguin contribution to DM direct detection.}
\label{fig:Zpenguin}
\end{figure}

The DM in this paper does not couple to the SM quarks and electron directly. 
However, an effective DM coupling to the quark vector currents is induced by a one-loop $Z$ penguin diagram, 
as shown in Fig.~\ref{fig:Zpenguin} 
in contrast to the DM study based on effective field theory with flavor changing mediators. 
The relevant interaction Lagrangian is written as follows~:
\begin{align}
\label{eq:Leff}
   \mathcal{L}_{\Sigma q}^{\rm eff}
   = \sum_{q = u, d, s} C_{V,q} \left\{ \Sigma^* i \partial_\mu \Sigma - (i \partial_\mu \Sigma^*) \Sigma \right\} (\bar{q} \gamma^\mu q)~.
\end{align}
The coefficient in Eq.~\eqref{eq:Leff} is given by
\begin{align}
    C_{V,q} &=
    a_Z \frac{1}{m_Z^2} \frac{g}{2 \cos \theta_W^{}} (T_3 - 2 Q_q \sin^2 \theta_W^{})~, \\
    a_Z &=
    \frac{(\kappa v)^2}{(4 \pi)^2} \frac{g}{2 \cos \theta_W^{}} \times \frac{1}{m_\rho^2 - m_\eta^2} \left[ f(m_\rho/m_\Sigma) - f(m_\eta/m_\Sigma) \right]
\end{align}
with $m_Z^{}$ being the $Z$ boson mass, $g$ the SU(2)$_L$ gauge coupling constant, $\theta_W^{}$ the Weinberg angle, $Q_q$ the electric charge of the SM quark $q$, and $T_3$ the third component of the weak isospin.
The loop function $f(x)$ is given by
\begin{align}
    f(x) =
    x^2 + x^2 (2 - x^2) \log x + x^3 \sqrt{x^2 - 4} \log \left( \frac{x + \sqrt{x^2 - 4}}{2} \right)~.
\end{align}
Given the above effective interaction, the spin-independent (SI) elastic scattering cross section between DM and a target nucleus with atomic (mass) number, $Z (A)$, is obtained by 
\begin{align}
\label{eq:xsec-scat}
   \sigma_{{\rm SI}}^{} =
   \frac{\mu_N^2 \left[ Z C_{V,p} +(A-Z) C_{V,n} \right]^2}{\pi} ~,
\end{align}
where $\mu_N^{}=m_\Sigma^{}m_N^{}/(m_\Sigma^{}+m_N^{})$ with $m_N^{}$ being the target nucleus mass and 
\begin{align}
   C_{V,p} = 2 C_{V,u} + C_{V,d}~,~~
   C_{V,n} = C_{V,u} + 2 C_{V,d}~. 
\end{align}

\subsection{Indirect Detection}
\label{subsec:indirect}

Fermi-LAT collaboration~\cite{Fermi-LAT:2009ihh} has observed gamma-ray flux from various dwarf galaxies, 
and their results have set upper limits on extra gamma-ray sources, for example, DM annihilation in such galaxies~\cite{Hoof:2018hyn}.
The bounds provided by Fermi-LAT are given for the lepton flavor conserving annihilation processes, such as $\Sigma \Sigma \to \mu \bar{\mu}, \tau \bar{\tau}$.
On the other hand, the DM in this paper annihilates into $\mu \bar{\tau}$ and $\tau \bar{\mu}$. 
Because one muon and one tau lepton are produced in each DM annihilation, the following quantity can be defined as an equivalent cross section of the lepton flavor conserving annihilation
\begin{align}
   (\sigma v_{\rm rel})_{\Sigma \Sigma \to \mu \bar{\mu}, \tau \bar{\tau}} \equiv
   \frac{(\sigma v_{\rm rel})_{\Sigma \Sigma \to \mu \bar{\tau}} + (\sigma v_{\rm rel})_{\Sigma \Sigma \to \tau \bar{\mu}}}{2}~.
\end{align}
Then, we compare 
$((\sigma v_{\rm rel})_{\Sigma \Sigma \to \mu \bar{\mu}, \tau \bar{\tau}} +  (\sigma v_{\rm rel})_{\Sigma^* \Sigma^* \to \mu \bar{\mu}, \tau \bar{\tau}})/4$ 
with the public cross section limits in deriving the Fermi-LAT bound on this model. 
The factor of 1/4 results from the fact that,  
assuming the symmetric relic, 
the number density of $\Sigma$ is half of the observed DM density, i.e. $n_\Sigma^{} = n_{\Sigma^*}^{} = n_\DM^{}/2$. 
The resulting $\mu, \tau$ flux from $\Sigma$ annihilation is smaller by 1/4 than the one from annihilation of self-conjugate DM, which is assumed in the Fermi-LAT paper.

DM also annihilates into heavy scalars and gauge bosons, such as $\phi^+ \phi^-, \rho \rho, \rho Z, \phi^- W^+$, if kinematically allowed. 
The daughter particles promptly decay into the SM fermions and contribute as additional gamma-ray source.
The spectrum of gamma rays emitted from such annihilations has a different behavior from that from $\Sigma \Sigma \to \mu \bar{\tau}, \tau \bar{\mu}$. 
It also depends on the polarization fraction of the daughter particles in each annihilation mode~\cite{Garcia-Cely:2016pse}. 
The resulting gamma-ray signatures can be an interesting prediction in our model. 
However, there is no applicable cross section limit for this annihilation channel in the market. 
An dedicated spectral analysis will be required to obtain the limit, but it is beyond the scope 
of this paper. We leave this study for future work.

\section{Constraints on Mediators}
\label{sec:mediator}

The mediators $\rho, \eta$ and its partner $\phi^\pm$ originate in the weak doublet scalar. 
The doublet directly couples to the SM gauge bosons via the electroweak interaction 
as well as $\mu$ and $\tau$ leptons via the flavor off-diagonal Yukawa interactions.
Through those interactions, various experimental constraints are imposed on the mediators.

\subsection{Precision measurements, triviality and potential stability}
\label{sec:precision}

The magnitude of $\mu\tau$ Yukawa interactions is not small in order to address the muon $g-2$ anomaly, 
so the processes associated with the muon and tau flavors can deviate from the SM prediction. 
In addition, the electroweak interactions of the extra scalars produce corrections to the vacuum polarization of the weak bosons 
whose effects are rendered in the $S,T,U$ parameters~\footnote{
A large correction to the $T$ parameter has recently been discussed in \cite{Han:2022juu} 
in relation to the new CDF measurement of the $W$ boson mass~\cite{CDF:2022hxs}. 
It seems that there is room to accommodate both the $W$ mass and muon $g-2$ 
by an additional contribution to the $T$ parameter from the extra scalars. 
We do not, however, pursue such a possibility in this paper and will refer to the PDG value of the $W$ boson mass~\cite{ParticleDataGroup:2020ssz}.
}. 
These observables are strongly constrained by precise measurements in the electroweak and lepton sectors. 
Effects of the $\mu\tau$-philic scalars on the electroweak precision tests, lepton universality, and triviality bound are examined in Refs.~\cite{Abe:2019bkf,Wang:2019ngf}. 
We do not repeat the detailed discussion here and leave them to the previous works,
while we shortly recapitulate the primary constraints: 
\begin{itemize}
\item Electroweak precision tests (EWPTs): the extra scalar contributions to the $S,T,U$ parameters have to be consistent with the SM. We evaluate $\chi^2$ with respect to $S=0.00\pm0.07$ and $T=0.05\pm0.06$~\cite{ParticleDataGroup:2020ssz} with $U=0$ fixed and require that $\Delta\chi^2=\chi^2-\chi^2_{\rm SM}$ are within $2\sigma$. 

\item Triviality bound: the quartic couplings and Yukawa couplings evolve with renormalization group equations (RGEs). We calculate the running couplings from the $Z$ boson mass scale at one-loop level and require $|\lambda_i| \leq 4\pi$ $(i=1\,$--$\,5)$, with $\lambda_1$ being the quartic coupling of the SM Higgs doublet, and $|y_t|, |y_{\mu\tau}|, |y_{\tau\mu}| \leq \sqrt{4\pi}$ up to 100\,TeV.

\item Potential stability: we require the RG-evolved couplings to satisfy all of the conditions $\lambda_1> 0, \  \lambda_2 > 0, \ 2\sqrt{\lambda_1\lambda_2}+\lambda_3>0, \ 2\sqrt{\lambda_1\lambda_2}+\lambda_3+\lambda_4-|\lambda_5|>0$~\cite{Kanemura:1999xf} at 100\,TeV~.
\end{itemize}

\subsection{Collider bounds}

There are also direct search constraints on the additional scalar doublet at the high-energy colliders. 
The searches for charged Higgs bosons by LEP~\cite{ALEPH:2013htx} gives a lower limit on the $\phi^\pm$ mass: $m_{\phi^\pm} \gtrsim 93.5\GeV$.
Moreover, the charged scalar has the same quantum charges as the left-handed charged sleptons in supersymmetric models except for the matter parity.
The slepton bounds from the electroweak productions, therefore, can be applied for $\phi^\pm$ when $\phi^+ \to \bar{\mu} \nu_\tau, \bar{\tau} \nu_\mu$ are dominant (prompt) decay processes.
For the case that $\phi^+ \to \bar{\mu} \nu_\tau$ is dominant, the bound on $\phi^\pm$ can be read from that of the left-handed smuon: $m_{\phi^\pm} \gtrsim 550\GeV$~\cite{ATLAS:2019lff}.
For the case that $\phi^+ \to \bar{\tau} \nu_\mu$ is dominant, the bound on the left-handed stau can be applied: $m_{\phi^\pm} \gtrsim 350\GeV$~\cite{CMS:2019eln,ATLAS:2019gti,CMS:2021woq}.
These bounds are not valid when $m_{\phi^\pm} \geq m_\rho + m_W^{}$ and the decay channel of $\phi^\pm \to \rho W^\pm$ is dominant.
In this case, the charged scalar can be lighter than $\sim 350\GeV$.
Note that there is a less constrained region for the stau-like case where $100\GeV \lesssim m_{\phi^\pm} \lesssim 115\GeV$ is still allowed even if the branching fraction of $\phi^+ \to \bar{\tau} \nu_\mu$ is 100\%~\cite{CMS:2021woq}.

Collider searches for the heavy neutral scalars with the $\mu\tau$ interactions have been studied in Refs.~\cite{Iguro:2019sly,Wang:2019ngf,Han:2021gfu,Wang:2021fkn}.
The signal processes are the electroweak production of the scalars such as $pp \to W^{\pm *} \to \phi^\pm \rho, \phi^\pm \eta$ and $pp \to Z^* \to \rho \eta$ followed by $\phi^\pm \to \tau^\pm \nu_\mu, \mu^\pm \nu_\tau$ and $\rho,\eta \to \tau^\pm\mu^\mp$.
The previous studies focus on the mass spectrum of $m_\rho < m_\eta \simeq m_{\phi^\pm}$ or $m_\eta < m_\rho \simeq m_{\phi^\pm}$, which are both different from ours, $m_{\phi^\pm}  \simeq m_\rho  < m_\eta$. 
Then, they point out that the LHC Run2 data will exclude $200\GeV \leq m_\eta \lesssim 500\GeV$~\cite{Iguro:2019sly,Wang:2019ngf}. 
They also explore the light mass region and show that $m_\rho < 20\GeV$ and $130\GeV < m_\eta \,(m_{\phi^\pm}) < 610\GeV$ are allowed~\cite{Wang:2021fkn}. 
These bounds are qualitatively the same in our case, but cannot be applied directly 
because of the different mass spectrum considered. 
In addition, in our case, $\rho$ and $\eta$ decay invisibly into a DM pair, which reduces the signal cross section. 
Thus, the bounds on the neutral scalars can be relaxed significantly. 

Given the above-mentioned collider bounds, we will later focus on two mass spectra for the extra scalars in our DM analysis: 
(i) $(m_{\phi^\pm}, m_\rho, m_\eta) = (100\GeV, 100\GeV, 130\GeV)$ and 
(ii) $(m_{\phi^\pm}, m_\rho, m_\eta) = (700\GeV, 680\GeV, 730\GeV)$.
In the spectrum (i), the extra scalars are light, and the charged scalar mass is placed on a blind spot in the current stau search. 
The viability of this mass window depends on the branching fraction of the neutral scalars to DM. 
There is a correlation between the invisible branching fraction and DM thermal production. 
We will explicitly show the correlation in Sec.~\ref{sec:result}. 
In the spectrum (ii), the collider bounds are certainly evaded, and the precision measurements and triviality bound provide a leading constraint.
In that sense, this is a relatively safe spectrum to examine the DM physics.

\subsection{Summary of the mediator bounds}

\begin{figure}[t]
\begin{minipage}[b]{0.49\linewidth}
\centering
\includegraphics[clip=true, scale=0.49]{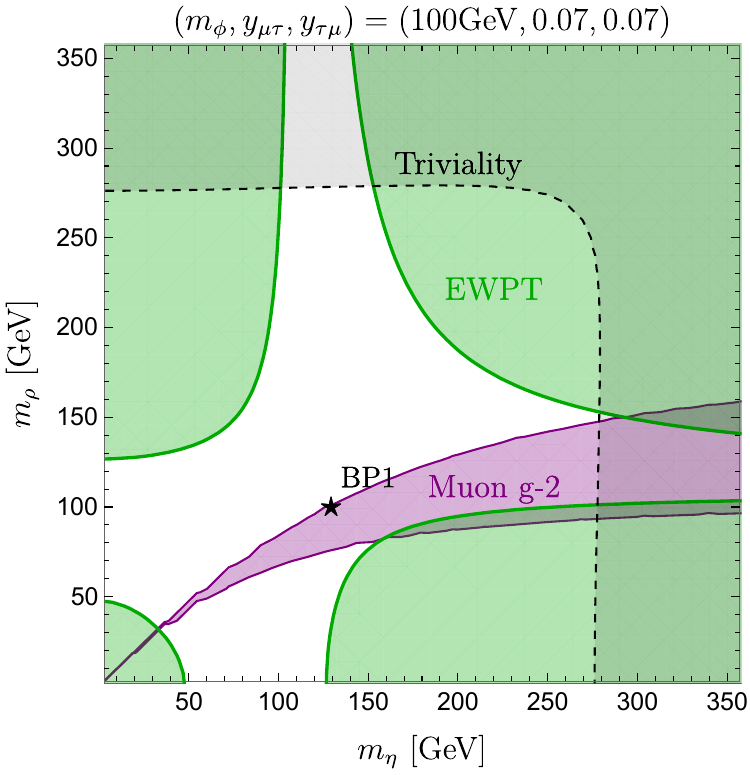}
\subcaption{BP1}
\end{minipage}
\begin{minipage}[b]{0.49\linewidth}
\centering
\includegraphics[clip=true, scale=0.49]{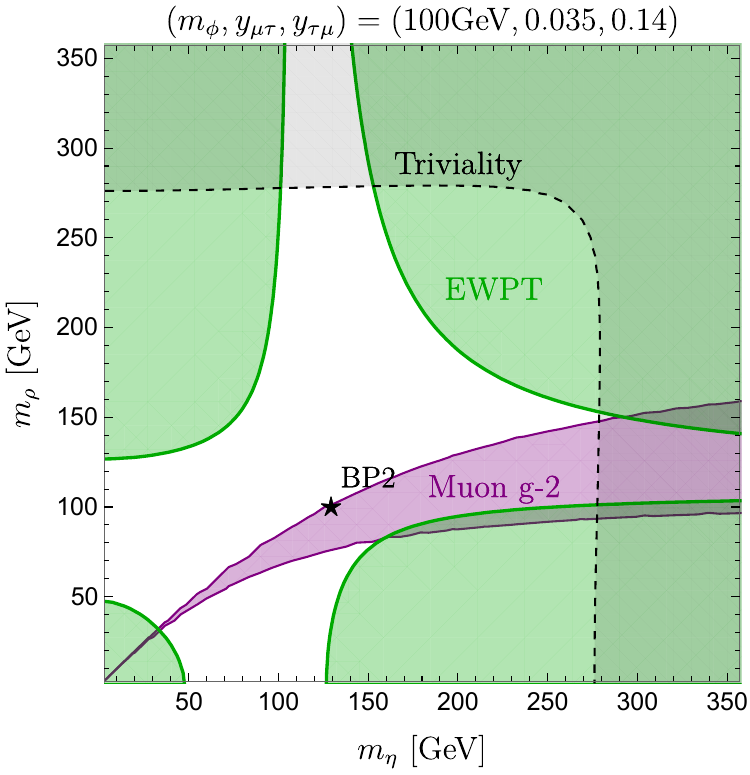}
\subcaption{BP2}
\end{minipage}\\
\begin{minipage}[b]{0.49\linewidth}
\centering
\includegraphics[clip=true, scale=0.49]{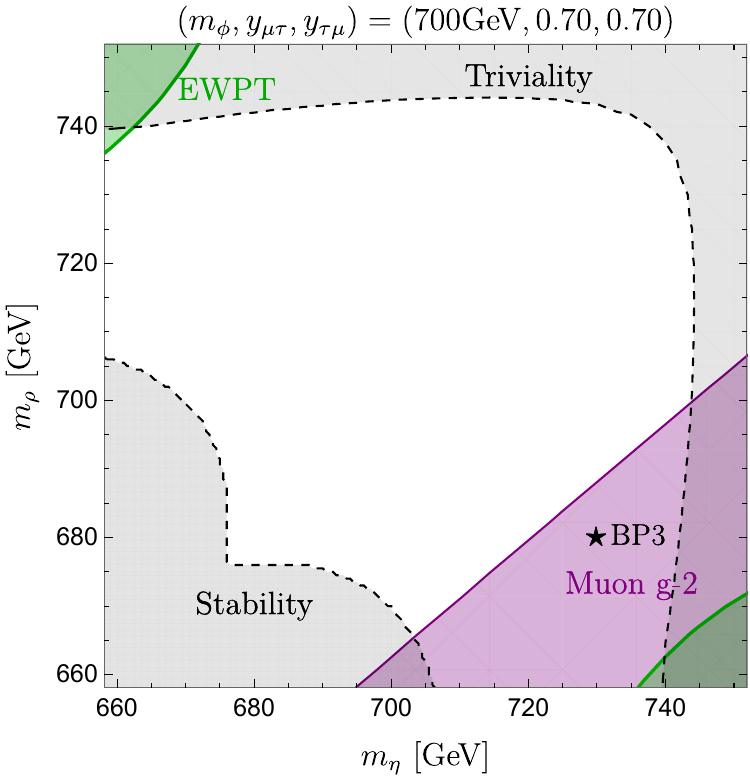}
\subcaption{BP3}
\end{minipage}
\begin{minipage}[b]{0.49\linewidth}
\centering
\includegraphics[clip=true, scale=0.49]{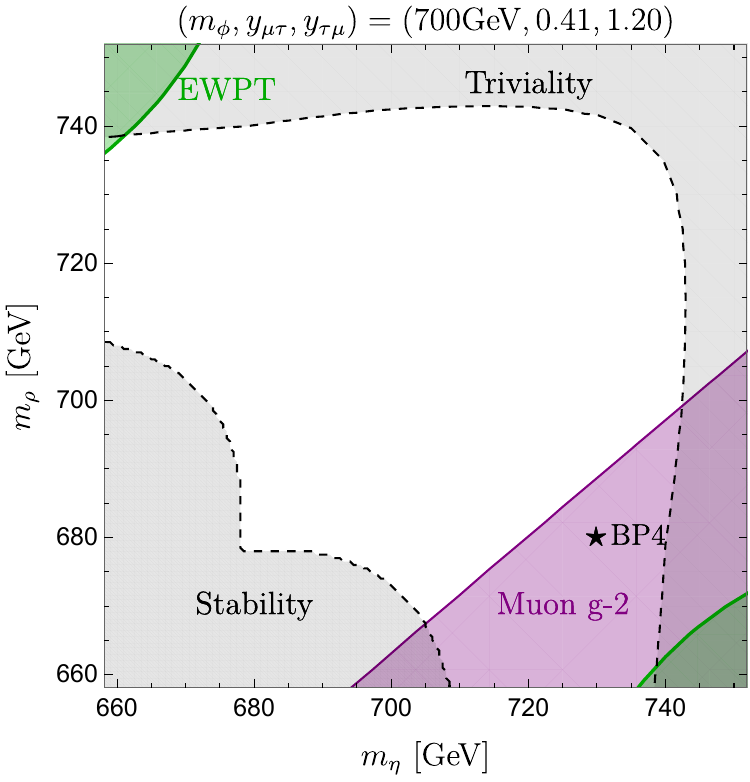}
\subcaption{BP4}
\end{minipage}
\caption{
Constraints on the $\mu\tau$-philic mediator 
with the four benchmark values of the charged scalar mass and flavor off-diagonal lepton Yukawa couplings.
The light gray (green) shaded region is excluded by the potential stability condition and triviality bound (EWPTs).
The purple shaded region is favored by the muon $g-2$.
The black star in each panel corresponds to the benchmark point used in our DM study in Sec.~\ref{sec:result}.
}
\label{fig:constraint-mediator}
\end{figure}

In Fig.~\ref{fig:constraint-mediator}, we show the bounds on the $\mu\tau$-philic mediators.
We pick four sets of values for the charged scalar mass and flavor off-diagonal lepton Yukawa couplings. 
In particular, the charged scalar mass is fixed to 100\,GeV (upper panels) and 700\,GeV (lower panels), inspired by the spectrum (i) and (ii) respectively. 
We also fix $\lambda_2=\lambda_3=0.5$ in all panels. 
The values of these couplings only affect the triviality and potential stability bounds. 

The light gray (green) shaded region is excluded by the potential stability condition and triviality bound (EWPTs).
The muon $g-2$ is explained in the purple shaded region within $2\sigma$.
It follows from Fig.~\ref{fig:constraint-mediator} that for the light charged scalar case (upper panels), 
the EWPTs provide the leading bound on the $\rho$ and $\eta$ masses below 270\,GeV. 
At least one of these neutral scalars should be degenerate in mass with the charged scalar. 
As their masses exceed 270\,GeV, large quartic couplings are required in order to generate 
appropriate mass spectrum. 
Thus the triviality bound is superior to the constraint from the precision tests. 
For the heavier case (lower panels), on the other hand, the potential stability condition and 
the triviality bound give the leading constraints.
Note that Fig.~\ref{fig:constraint-mediator} is similar to Fig.~5 of Ref.~\cite{Abe:2019bkf} except the horizontal and vertical axes are the masses of $\eta$ and $\rho$, instead of the quartic couplings, $\lambda_4$ and $\lambda_5$.
Our plots are thus obtained approximately by rotating the plots in the literature by 45 degrees counterclockwise.~\footnote{
The triviality and potential stability bounds in this paper are different from those of Ref.~\cite{Abe:2019bkf} since there are some errors in the $\beta$ functions for the RGEs in Appendix A of Ref.~\cite{Abe:2019bkf}. 
The $\beta$ functions in Ref.~\cite{Abe:2019bkf} paper agree with those in Ref.~\cite{Goudelis:2013uca} by replacement as $(\lambda_3,\lambda_4) \to (\lambda_3+\lambda_4,-\lambda_4)$. 
}

\section{Result}
\label{sec:result}

For references, we choose four benchmark points (BPs), marked with the black star in Fig.~\ref{fig:constraint-mediator}, which can all accommodate the muon $g-2$ anomaly and pass the various theoretical and experimental constraints~: 
\begin{align*}
\label{eq:benchmark}
   &{\rm BP}1~:~ 
   (m_{\phi^\pm}, m_\rho, m_\eta) = (100\GeV, 100\GeV, 130\GeV),\quad (y_{\mu\tau}, y_{\tau\mu}) = (0.07, 0.07)~, \\
   &{\rm BP}2~:~ 
   (m_{\phi^\pm}, m_\rho, m_\eta) = (100\GeV, 100\GeV, 130\GeV),\quad (y_{\mu\tau}, y_{\tau\mu}) = (0.035, 0.14)~, \\
   &{\rm BP}3~:~ 
   (m_{\phi^\pm}, m_\rho, m_\eta) = (700\GeV, 680\GeV, 730\GeV),\quad (y_{\mu\tau}, y_{\tau\mu}) = (0.7, 0.7)~, \\
   &{\rm BP}4~:~ 
   (m_{\phi^\pm}, m_\rho, m_\eta) = (700\GeV, 680\GeV, 730\GeV),\quad (y_{\mu\tau}, y_{\tau\mu}) = (0.41, 1.2)~. 
\end{align*}

\begin{figure}[t]
\begin{minipage}[b]{0.49\linewidth}
\centering
\includegraphics[clip=true, scale=0.49]{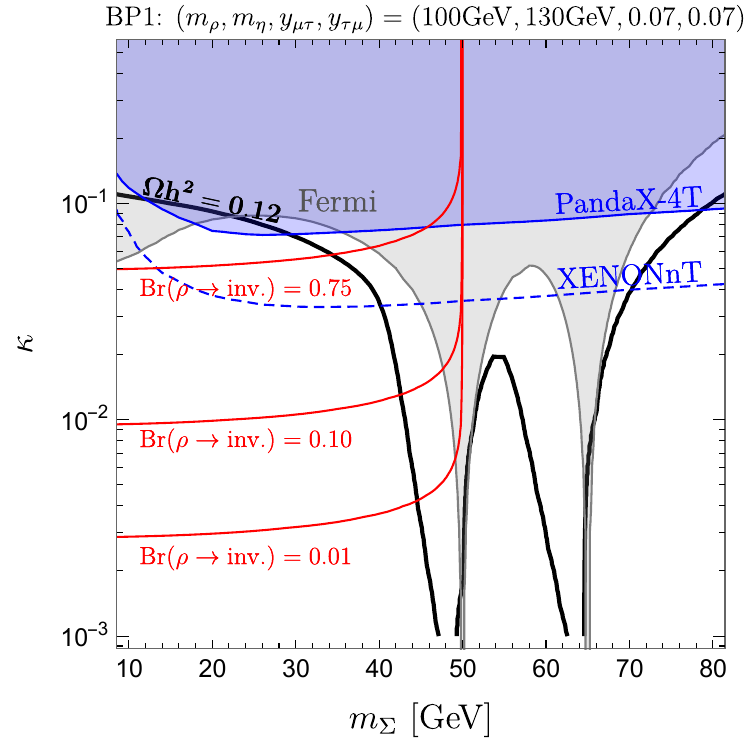}
\subcaption{BP1}
\end{minipage}
\begin{minipage}[b]{0.49\linewidth}
\centering
\includegraphics[clip=true, scale=0.49]{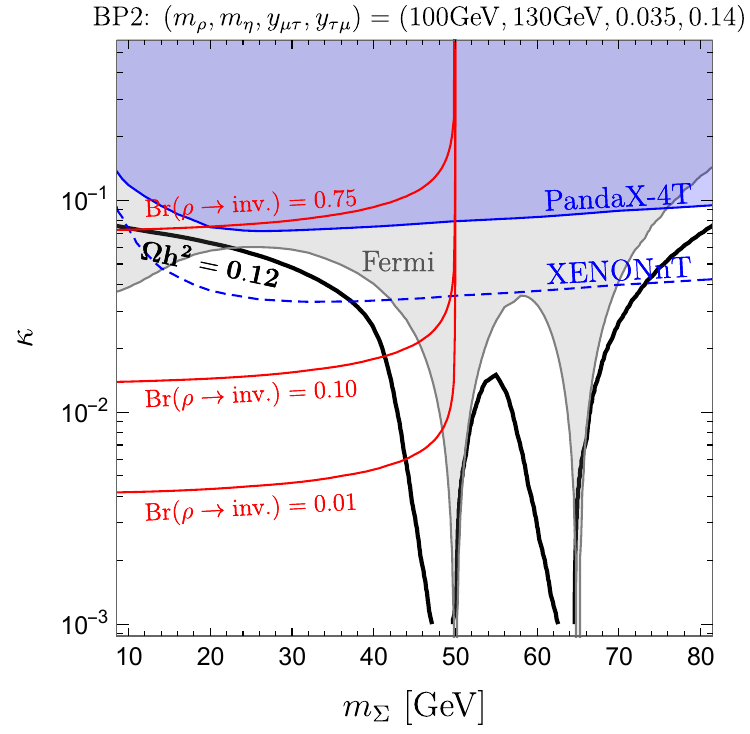}
\subcaption{BP2}
\end{minipage}\\
\begin{minipage}[b]{0.49\linewidth}
\centering
\includegraphics[clip=true, scale=0.49]{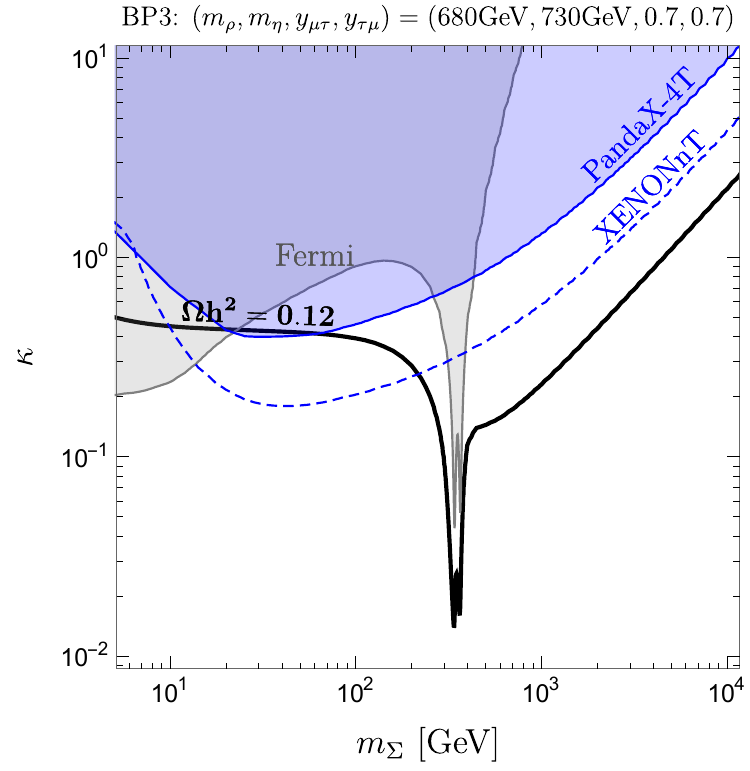}
\subcaption{BP3}
\end{minipage}
\begin{minipage}[b]{0.49\linewidth}
\centering
\includegraphics[clip=true, scale=0.49]{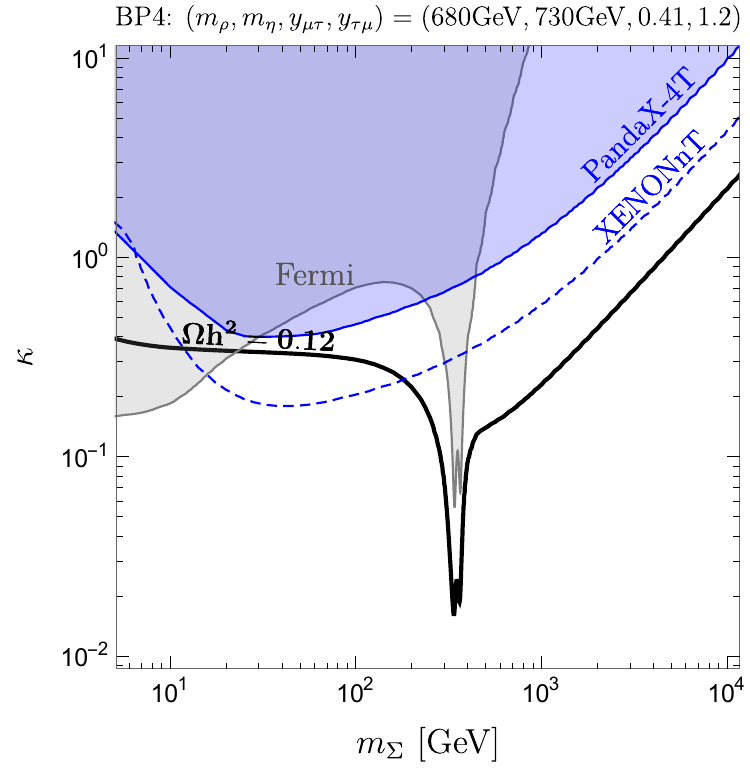}
\subcaption{BP4}
\end{minipage}
\caption{
Experimental upper limits on the scalar quartic coupling $\kappa$ for the BP1 (top-left), BP2 (top-right), BP3 (bottom-left), and BP4 (bottom-right).
The blue and gray shaded regions are excluded by PandaX-4T experiment~\cite{PandaX-4T:2021bab} and Fermi-LAT observation~\cite{Hoof:2018hyn}.
The blue dashed curve represents the future prospect of the XENONnT experiment~\cite{XENON:2020kmp}.
In the top two panels, the branching fractions of $\rho \to {\rm inv.}$ are shown by the red solid curves.
}
\label{fig:constraint-DM}
\end{figure}

Figure \ref{fig:constraint-DM} shows the constraints on the DM for the above four benchmark parameter sets.
The black solid curve corresponds to $\Omega_{\rm DM}^{} h^2 = 0.12$, and the region below this line is excluded by DM overabundance.
We also show the excluded regions by direct detection via the $Z$ penguin diagram (Fig.~\ref{fig:Zpenguin}) and indirect one via annihilation into $\mu \bar{\tau}$ and $\tau \bar{\mu}$.
The blue and gray shaded regions are excluded by PandaX-4T experiment~\cite{PandaX-4T:2021bab} and  Fermi-LAT observation~\cite{Hoof:2018hyn}, respectively, while the blue dashed curve represents the future sensitivity at the XENONnT experiment~\cite{XENON:2020kmp}.
The DM annihilation into charged leptons is also constrained by the spectrum distortion of cosmic microwave background (CMB)~\cite{Slatyer:2015jla,Leane:2018kjk}. The CMB bound is, however, weaker than the Fermi bound in the parameter space of our interest, so that we omit the former one in Fig.~\ref{fig:constraint-DM}.
The branching fractions of $\rho$'s invisible decay are shown by the red solid curves in the top-left (BP1) and top-right (BP2) panels as references.

One can see two common features to every panel as for the DM constraints.  
The first one is that the contours of the DM relic abundance and the exclusion limits of the indirect detection have two steep‐walled valleys, which are caused by resonant annihilation of the DM at $m_\Sigma^{}\simeq m_\rho/2$ and $m_\eta/2$.
The second one is that the Fermi-LAT bound excludes the DM with its mass below 20\,GeV if the observed DM abundance is explained by the thermal relics of $\Sigma$ and $\Sigma^*$.

For the light DM and lepton flavor changing mediator case (BP1 and BP2), 
the current direct detection constraints are avoided in $20\GeV \lesssim m_\Sigma^{}\lesssim 80\GeV$. 
Future XENONnT experiment can further probe $m_\Sigma^{}\simeq 10\,$--$\,40\GeV$ and $70\,$--$\,80\GeV$.
We remind that these two BPs may be excluded by the LHC data unless $\rho$ and $\eta$ decay mostly to DM. 
Thus the mass region heavier than $m_\eta/2$ would be ruled out, although no experimental search 
for this mass region is found to our best knowledge. 
When DM is lighter than $m_\eta/2$, the branching fractions of $\rho$ and $\eta$ are crucial. 
It follows from Fig.~\ref{fig:constraint-DM} that the invisible branching fraction of 
$\rho$ can be larger than 80\% for BP1 and 60\% for BP2.
This means that the signal cross section can be reduced by a corresponding factor. 
To explore this parameter space, however, a dedicated collider analysis is desired. 

For the heavy DM and lepton flavor changing mediator cases (BP3 and BP4), 
the latest PandaX result excludes the DM with $m_\Sigma^{}\lesssim 60\GeV$. 
The mass region of $m_\Sigma^{}\simeq 60\,$--$\,200\GeV$ is within the reach of the future XENONnT experiment.
Otherwise, this DM candidate is free from the current and future planned experiments.

\section{Summary}
\label{sec:summary}

We have studied a simple scalar DM model based on the $Z_4$ lepton flavor symmetry.
In this model, the scalar singlet DM, scalar doublet mediator, muon, and tau lepton are nontrivially transformed under the $Z_4$ transformation.
The DM interacts with the SM fields through the scalar mediator, which has flavor off-diagonal couplings to the muons and tau leptons. 
The observed relic abundance of the DM is realized by the thermal freeze-out mechanism.
The scalar mediators only couple to the muon and tau lepton at the tree level, 
while have no interaction with the quarks and electrons.
Therefore, the model can explain the DM relic density and the muon $g-2$ anomaly simultaneously 
without conflicting the current severe constraints by various experiments and observations.
We have evaluated direct and indirect bounds on the DM.
Because the $\mu\tau\mathchar`-$philic mediator in our model comes from the SU(2)$_L$ doublet, the one-loop $Z$ penguin diagram gives a significant contribution to the SI elastic scattering between the DM and nucleon.
The allowed mass ranges of the DM can be further explored by the direct detection experiment in the future 
except for the resonant and heavy DM mass region. 
In this paper, we focus on the $\mu\tau\mathchar`-$philic scalar mediator, and on the other hand, the $e\mu$ and $e\tau\mathchar`-$philic cases induce interesting phenomena, for example, the DM can scatter inelastically with electrons in material, and muon and tau lepton are produced as recoil particles.
This will be shown elsewhere.

\section*{Acknowledgments}
This work is supported in part by the Grant-in-Aid for Research Activity Start-up (No.21K20365 [KA]) and Scientific Research (No.18H01210 [KA], No.22K03620 [KT]) and MEXT KAKENHI Grant (No.18H05543 [KA and KT]).
S.O. acknowledges financial support from the State Agency for Research of the Spanish Ministry of Science and Innovation through the ``Unit of Excellence Mar\'ia de Maeztu 2020-2023'' award to the Institute of Cosmos Sciences (CEX2019-000918-M) and from PID2019-105614GB-C21 and 2017-SGR-929 grants.

\appendix
\section{Neutrino masses and mixing}
\label{app:neutrino}

In this appendix, we illustrate an extension of the model in Sec.~\ref{sec:model} for the realization of the neutrino mass and mixing.
We additionally introduce three right-handed neutrinos $(N_{e}, N_{\mu}, N_{\tau})$ and a $Z_4$-breaking singlet scalar $S$ to the model in Sec.~\ref{sec:model}.
In this model, the $Z_4$-charged scalar $S$ acquires a nonzero VEV and breaks the $Z_4$ lepton flavor symmetry.
Therefore, the dark parity ($Z_2$ symmetry) is introduced in order to keep the DM stability.
Quantum numbers of this extended model under the $Z_4 \times Z_2$ symmetry are listed in Tab.~\ref{tab:fields-Z4Z2}.
\begin{table}[tb]
\centering
\begin{tabular}{|c|ccccc:cc|}
\hline
particles & $(L_{e}, L_{\mu}, L_{\tau})$ & $(e_{R}^{}, \mu_{R}^{}, \tau_{R}^{})$ & $H$ & $\Phi$ 
& $\Sigma$ & $(N_{e}, N_{\mu}, N_{\tau})$ & $S$ \\ 
\hline 
SM & $(\bm{1},\bm{2})_{-1/2}$ & $(\bm{1},\bm{1})_{-1}$ & $(\bm{1},\bm{2})_{1/2}$ & $(\bm{1},\bm{2})_{1/2}$ & $(\bm{1},\bm{1})_{0}$ & $(\bm{1},\bm{1})_{0}$ & $(\bm{1},\bm{1})_{0}$ \\
$Z_{4}$ & $(1, i, -i)$ & $(1, i, -i)$ & $1$ & $-1$ 
& $i$ & $(1, i, -i)$ & $i$ \\ 
$Z_{2}$ & $+$ & $+$ & $+$ & $+$ & $-$ & $+$ & $+$ \\ 
\hline 
\end{tabular}
\caption{
The particle content of $Z_{4}\times Z_2$ symmetric model.
The quantum numbers of the SM are also shown in the notation of $({\rm SU(3)}_C, {\rm SU(2)}_L)_{{\rm U(1)}_Y}$.
}
\label{tab:fields-Z4Z2}
\end{table}

Under the $Z_{4}\times Z_2$ symmetry, the Lagrangian for the neutrino mass generation sector is written by
\begin{align}
   {\mathcal L}_{N} &= 
   -\frac{1}{2} \begin{pmatrix} \overline{N_{e}^{c}} & \overline{N_{\mu}^{c}} & \overline{N_{\tau}^{c}} \end{pmatrix} 
   \begin{pmatrix} M_{ee} & \lambda_{e\mu} S^{*} & \lambda_{e\tau}S \\ \lambda_{e\mu} S^{*} & & M_{\mu \tau} \\ \lambda_{e\tau} S & M_{\mu \tau} & \end{pmatrix} 
   \begin{pmatrix} N_{e}^{} \\ N_{\mu}^{} \\ N_{\tau}^{} \end{pmatrix} \nonumber \\
   & \qquad 
   - \begin{pmatrix} \overline{L_{e}} & \overline{L_{\mu}} & \overline{L_{\tau}} \end{pmatrix} 
   \begin{pmatrix} y_{ee}^{} \tilde{H} & & \\ & y_{\mu \mu}^{} \tilde{H} & y_{\mu \tau} \tilde{\Phi} \\ & y_{\tau \mu} \tilde{\Phi} & y_{\tau \tau}^{} \tilde{H} \end{pmatrix} 
   \begin{pmatrix} N_{e}^{} \\ N_{\mu}^{} \\ N_{\tau}^{} \end{pmatrix} +\text{H.c.}~,
\end{align}
where $\tilde{H} = i \sigma_2 H^*,\, \tilde{\Phi} = i \sigma_2 \Phi^*$ with $\sigma_2$ being the second Pauli matrix.
After acquiring the VEV of $H$ and $S$, the neutrinos obtain the Dirac and Majorana mass terms, respectively.
If the Majorana masses are much heavier than the Dirac ones, the light masses of the active neutrinos is naturally realized by the seesaw mechanism~\cite{Minkowski:1977sc,Yanagida:1979as,Gell-Mann:1979vob,Mohapatra:1979ia}.

Because of the remnant of the $Z_4$ lepton flavor symmetry, the mass matrix of the light neutrinos has the so-called two-zero minor  structure~\cite{Araki:2012ip,Crivellin:2015lwa,Asai:2017ryy,Asai:2018ocx,Asai:2019ciz,Asai:2020qax}. With this structure the lightest neutrino mass and CP phases are determined as functions of the neutrino oscillation parameters, such as the mixing angles and mass squared differences.
According to Refs.~\cite{Asai:2017ryy,Asai:2018ocx,Asai:2019ciz,Asai:2020qax}, the extended model in this appendix realizes the observed neutrino mixing at $3\sigma$ level without conflicting with the neutrino mass bound from the CMB measurement~\cite{RoyChoudhury:2019hls}. 

{\small
\bibliography{ref}
}

\end{document}